# Thinking about Archeoastronomy

Noah Brosch

The Wise Observatory and the Raymond and Beverly Sackler School of Physics and Astronomy, Tel Aviv University, Tel Aviv 69972, Israel

*Abstract*

I discuss various aspects of archeoastronomy concentrating on physical artifacts (i.e., not including ethno-archeoastronomy) focusing on the period that ended about 2000 years ago. I present examples of artifacts interpreted as showing the interest of humankind in understanding celestial phenomena and using these to synchronize calendars and predict future celestial and terrestrial events. I stress the difficulty of identifying with a high degree of confidence that these artifacts do indeed pertain to astronomy and caution against the over-interpretation of the finds as definite evidence.

With these in mind, I point to artifacts that seem to indicate a human fascination with megalithic stone circles and megalithic alignments starting from at least 11000 BCE, and to other items presented as evidence for Neolithic astronomical interests dating to even 20000 BCE or even before. I discuss the geographical and temporal spread of megalithic sites associated with astronomical interpretations searching for synchronicity or for a possible single point of origin.

A survey of a variety of artifacts indicates that the astronomical development in antiquity did not happen simultaneously at different locations, but may be traced to megalithic stone circles and other megalithic structures with possible astronomical connections originating in the Middle East, specifically in the Fertile Crescent area. The effort of ancient societies to erect these astronomical megalithic sites and to maintain a corpus of astronomy experts does not appear excessive.

*Introduction*

This paper deals with "archeoastronomy" in a restricted sense. Before going deeper into the subject it is worth thinking a bit about this term, which is a combination of two words; what do we mean by archeoastronomy? One definition, found in the Wikipedia, is "the study of how past people have understood the phenomena in the sky, how they used phenomena in the sky and what role the sky played in their cultures". Another one, from the web page of the Center for Archeoastronomy at the University of Maryland[1], is the "study of the astronomical practices, celestial lore, mythologies, religions and world-views of all ancient culture". Clive Ruggles (Emeritus Professor of archaeoastronomy at the University of Leicester) defines Archeoastronomy as "the study of beliefs and practices relating to the sky in the past, especially in prehistory, and the uses to which people's knowledge of the skies was put"[2]. Here I restrict the definition of archeoastronomy to relate to any material artifact that can be associated with celestial objects and the discussion to be limited to the period ending some 2000 years ago. An artifact is, in an archeological sense, an object made or used by humans. I exclude from this discussion various written documents that have obvious astronomical connections (e.g., the Babylonian astronomical records of the visibility of Venus appearing in tablet 63 of the series *Enuma Anu Enlil* that dates from the middle of the 17$^{th}$ century BCE, or the Babylonian list of stars or constellations called MUL.APIN and dated from the end of the 12$^{th}$ century BCE, both discussed and various versions compared in Hobson's thesis[3]).

Archeoastronomy in the sense described here has been pioneered by Thom[4] and later by Hoskin[5], among others. Thom strived to put archeoastronomy on a quantitative and statistically-significant sound base. Hoskin, who studied the orientation of Mediterranean temples and tombs, discussed in chapter 3 of his book the "tally stones" at the entrance of the Muajdra III temple in Malta, dated at about 3600 BCE (p. 31 et seq.). These were described also by Ventura et al.[6] who concluded that the holes in the tally stones recorded the number of days that passed between heliacal risings[7] of "significant" stars (e.g., Sirius) or asterisms (e.g., the Pleiades). These papers emphasize the major uncertainty of such interpretations: from a set of holes in a rock the investigators presented far-reaching conclusions about the intellectual and scientific capabilities in the ancient world.

---

[1] http://terpconnect.umd.edu/~tlaloc/archastro/.
[2] http://www.cliveruggles.net/.
[3] Hobson, R. 2009. *THE EXACT TRANSMISSION OF TEXTS IN THE FIRST MILLENNIUM B.C.E.* PhD thesis, University of Sydney.
[4] e.g. Thom, A. 1974. *Astronomical significance of prehistoric monuments in Western Europe*. Phil. Trans. R. soc. Lond. A **276**, 149-156.
[5] Hoskin, M. 2001. *Tombs, Temples and their orientation*. Ocarina Books, UK.
[6] Ventura, F., Serio, G.F. and Hoskin, M. 1993. *Possibly Tally Stones at Mnajdra, Malta*. JHA XXIV, 171-183.
[7] The heliacal rising of a celestial body happens when it briefly becomes visible on the eastern horizon just before sunrise, after a period when it was not visible.

The techniques used by Thom, Hosking and others require accurate measurements at a large number of sites. The measures parameters are directions of temple axes, of external walls, of viewing directions, or of marker stone alignments with respect to cardinal directions. The altitude of the local horizon for each location and each sight line is also measured relative to horizontal. The astronomical application comes when the azimuths and altitudes are matched with the rising or setting positions of a variety of celestial bodies (Sun, Moon, planets, bright stars, and prominent asterisms. All these parameters should carry with them a measurement error, yet not all cases in the literature take this trouble. In calculating the visibility of phenomena the following are taken into account: precession, refraction, and extinction. Normally, the proper motion of a body is not included in the calculation and neither are possible changes of the location and of the horizon due to tectonic movements and land erosion. Sometimes the investigator is not even aware of the limitations of human vision, an angular resolution of 1-2 arcminutes, and quotes the findings to absurd accuracies.

Within the restricted definition and limited time frame, my questions are:

(a) What kinds of artifacts have been associated with archeoastronomy?
(b) What dates can one assign to the various artifacts?
(c) How much effort was required to produce the artifact?
(d) Does an archeoastronomical interpretation of an artifact make sense?

From these questions and their answers one can attempt to address the even more interesting issue of whether one can argue for synchronicity, i.e., whether the apparition of astronomically-related artifacts took place in a single, relatively short period in a multitude of locations, or whether there was a slow diffusion of astronomical lore from one or a few culture centers to the rest of the sites (approximately the hyperdiffusionism theory). This question was previously discussed by e.g., Atkinson[8] in the geographical context of northern and western Europe.

To set the stage for approaching the questions posed above, I dwell briefly on the topic of the size of the human population throughout the ages. Various estimates put the size of the entire human population to only about three million at 9000 BCE[9], 10 million around 6000 BCE, and

---

[8] Atkinson, R.J.C. 1974. *II. Ancient Astronomy: Unwritten Evidence*. Phil. Trans. R. Soc. London A **276**, 123-131.
[9] I am using BCE to indicate dates "before the common era"; the first draft of this paper was written in 2010 CE.

25 million near 3000 BCE. The world population was probably 100 million individuals[10] when ancient Greece reached its intellectual glory, but the majority was concentrated in a small number of regions (Mediterranean rim and Nile hinterland, the Indus valley and Mesopotamia, the Far East)[11]. These population sizes, which undoubtedly carry large errors, are important because they determine the fraction of the population that could be "exempted" from the daily struggle of finding food in order to perform astronomical observations, or create/build astronomical observatories. The exempted fraction of the population would not only have to develop astronomical observation methods, but also keep records and find ways to convey the collected information to those who would carry the effort on after the first such specialized individuals would pass away, and derive algorithms to model the collected data. I will return to this topic later, in the interlude preceding the concluding third of the paper.

This paper is constructed as follows: I first discuss briefly time intervals of relevance to prehistoric man. I then bring a number of examples, first from the Mediterranean region, then from other parts of the globe, that point to an early interest in celestial phenomena. These examples, resulting from investigations conducted by the end of the 20th century and the beginning of the 21st, cause some doubts to be raised about the possible over-interpretation of the findings. I then bring examples of more recent studies where more rigorous as well as multi-disciplinary techniques were used. Finally, I present some personal conclusions about the results based on a partial and incomplete evaluation of this discipline from the point of view of an observational astronomer.

*The cycles of life*

We live a life of cycles. The daily alternation of light/dark and sunrise/sunset imposes the same cycle on physical mechanisms within our bodies (circadian rhythm). Not only humans exhibit this cycle, but also many other life forms, such as animals, plants, fungi and bacteria (eukaryotic and cyanobacteria). It is possible that the daily cycle of the internal clock appeared in the earliest single-cell organisms as a protection against damaging UV during DNA replication.

Another well-known human cycle is the 28-day menstrual one in human females. While some associated this with moonlight, with a claim for ovulation at full Moon and menstruation at new

---

[10] The world population estimates are from Wikipedia, extracted on 27 November 2010. Originally, they are from a compilation of sources by the US Census Bureau.

[11] The literature shows considerable controversy regarding the size of the population in the Americas during the period considered here.

Moon[12], a comparison with other primates indicates that this lunar connection is probably a coincidence, since the 28-day cycle of human and orang-utan females is different from the 30 days in gorillas and from the 35-day cycle of chimpanzees. Also, it is possible that in human societies the menstrual cycle synchronizes among females in close proximity (the McClintock effect[13]) and not with astronomical bodies.

In addition to the day-night and monthly cycles other rhythms influenced the life in the ancient world. An important one was the turning of the seasons that, in temperate and Polar Regions, is marked by changes in the sunlight intensity that cause hibernation or migration of animals; this definitely affected hunting and plant gathering opportunities that provided food for the early humans. In the Nile valley, the cyclical floods of the river upon the melting of snows in the Central African mountains and the monsoon rains at the sources of the Blue Nile (Lake Tama in Ethiopia) and the White Nile (Lake Victoria in Tanzania) were the main regulator of plant growth.

I will not discuss circadian cycles. The seasonal ones, however, have probably had a strong influence on the well-being and indeed on the survivability of ancient societies. The ability to predict natural phenomena, such as the beginning of the rainy season, or the expectations of a draught, certainly influenced the gathering of food and hunting of animals, and the agricultural production at a later time. This, without the buffer offered by long-term produce surplus and storage, could cause famine on one hand or abundance on the other. Remember the seven skinny cows and the seven fat ones in Pharaoh's dream[14]; this brings us naturally to the topic of Ancient Egypt.

*Ancient Egypt*
Before about 9000 BCE, the Saharan desert had no settlements beyond the Nile valley. During the 5th millennium BCE Egyptian Neolithic societies were entering the "early predynastic" phase, a transformation completed by different cultures in the region from the Nile Delta in the North to the Upper Nubia in the South by the end of the 5th millennium. These peoples were the result of the cultural and racial mixing of the local inhabitants of the Nile valley with migrants coming into the valley from the West (especially at southern latitudes) and from the East (along the Mediterranean coast line into the Nile Delta and the Fayum region).

---

[12] Knight, C., Power, C., Watts, I. 1995. *The Human Symbolic Revolution: A Darwinian Account*, Cambridge Archaeological Journal **5(1)**, 75-114.
[13] McClintock MK (1971). *Menstrual synchrony and suppression*. Natur*e* **229** (5282): 244-245.
[14] Genesis 41, 1-7.

The migrations from the Eastern Sahara and Western Egyptian oases were probably caused by the desertification of what was until then a savannah, since palaeoclimatology showed that the last wet phase in the Sahara occurred from about 8000 to 5500 BCE and ended with the "5.9 kilo-year event[15]". Similar climatic changes and related population movements already happened in later Middle Palaeolithic when the desert was first abandoned and then re-occupied in the early Holocene. The Nile Valley, with its steady and abundant water supply, was an obvious attraction point for migrating Saharan and Near Eastern inhabitants.

When settled in the Nile Valley around 6000 BCE, the migrants became agriculturalists under the influence of new immigrants into the same regions coming from the Fertile Crescent region that was affected by similar climate changes[16]. The proto-Egyptians established larger settlements, towns, local kingdoms, etc., and founded a new kind of economy and social structure that required the performance of synchronized activities. They developed a need of time measuring and time keeping. One of their earliest methods used an astronomical phenomenon to regulate the calendar and set the time of sowing based on observing the heliacal rising of Sirius (α Canis Majoris).

In ancient times, probably even before the First Dynasty (started by Pharaoh Menes/Narmer who reigned from ~3100 to ~3050 BCE), it was observed that the Nile floods happened a short time after the heliacal rising of Sirius. This star thus became the primary calibrator of the Egyptian calendar for two millennia, beginning from the First Dynasty. Since Sirius is the brightest star in the sky, it would have been relatively easy to see in the pre-dawn sky either with the help of natural or artificial sight-lines pointing to its rising position, as perhaps the Naqada III[17] Hathor palette shows, or via an approximately equilateral triangle formed with two other bright stars (Betelgeuse=α Orionis and Castor=α Gemini) that would have been well above the horizon with Sirius just below the horizon[18].

---

[15] An intense aridification event that probably initiated the drying out of the Sahara desert at about 3900 BCE. Coincides with the Bond 4 climatic event; see Bond, G. Showers, W., Cheseby, M., Lotti, R., Almasi, P., deMenocal, P., Priore, P., Cullen, H., Hajdas, I. Bonani, G. 1997. *A Pervasive Millenial-Scale Cycle in North Atlantic Holocene and Glacial Climates*. Science **278**, 1257-1266.

[16] Parker, A.G., Goudie, A.S., Stokes, S., White, K., Hodson, M.J., Manning, M., Kennet, D. 2006. *A record of Holocene climate change from lake geochemical analyses in southeastern Arabia*. Quaternary Res. **66**, 465-476.

[17] The name of a late predynastic=neolithic period, about 3300 BCE, some 200 years preceding Narmer's palette The Hathor palette is exhibited in the Petrie Museum of Egyptian Archaeology at University College London**.**

[18] Brosch, N. 2008. *Sirius Matters*. Springer. ISBN 1-4020-8318-1

It is likely that the people who noted the astronomical-economic coincidence and conducted such observations served a very important function in planning the activities of an agrarian society. These priest-astronomers wielded considerable power; they probably mixed the astronomical knowledge of when the flood waters that revive the vegetation are likely to arrive with mystical beliefs (e.g., the re-birth of Osiris) while checking via their Nilometers whether the flood indeed started. The importance of Sirius as a way to extrapolate the future Nile flood linked it with the goddess Hathor (depicted as cow-like, sometimes with a star between the horns).

The plateau of Giza was selected by Pharaoh Khufu, also called "Keops" in ancient sources, in the middle of the third millennium BCE (about 2550 BCE), as his final resting place. Khufu was the second pharaoh of the fourth Dynasty. His burial site includes a large pyramid, a large family necropolis and possibly the Sphinx (although there are claims that the Sphinx predated the pyramids). Part of the necropolis was completed by his son Khafre, who added several monuments and completed the second pyramid. Menkaure, Khafre's son, constructed the third pyramid and associated monuments at an even later date. The entire period of construction of the three large pyramids lasted some 80 years.

The three main Giza pyramids are oriented with very good precision, much better than one degree, to the cardinal directions. One possibility discussed in the literature is that this orientation accuracy, for such huge monuments, could only have been achieved by astronomical observation, possibly of the 'imperishable stars' [circumpolar stars] such as Meskhetyu (Ursa Major) that were circling the north celestial pole and were always above the horizon, never setting for observers in Egypt. Since at the time the pyramids were built no bright star coincided with the pole (the precession of the Earth's axis did not yet bring Polaris near the North Celestial Pole), it is interesting to speculate how the ancient Egyptians were finding north with such great accuracy.

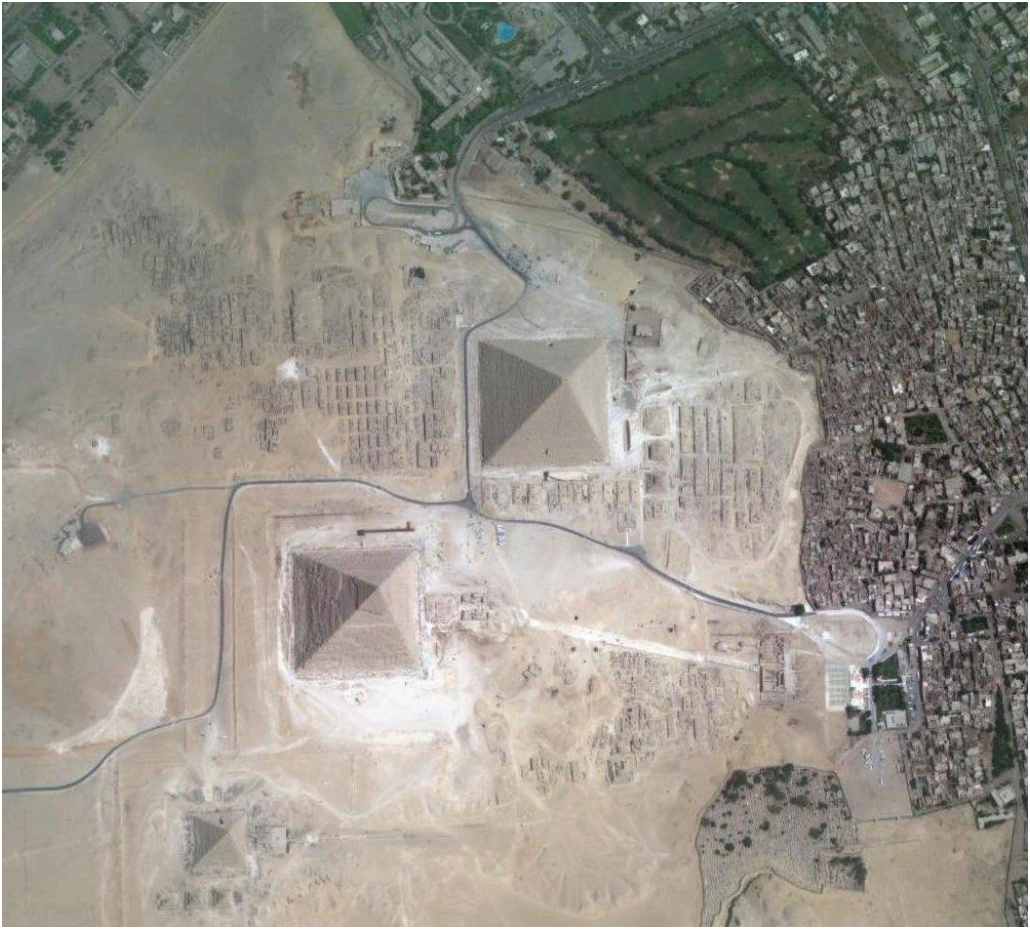

Figure 1: Giza plateau with the three large pyramids (from Google Earth). North at the top and East at the right.

Kate Spence[19] proposed that the method of orienting the large pyramids was by using stars to determine the direction to the North at night. The method involves forming a virtual line by connecting one star in Ursa Minor (Kochab=β UMi) with another star in Ursa Major (Mizar= ζ UMa) and waiting until this virtual line is exactly vertical (as measured e.g., with a plumb line). As evidence in support of this hypothesis she poined to the proposed star-connecting line precession past the north celestial pole at a rate of 27 arcminutes per century[20]. This alignment could therefore be used to find north only for a relatively short period around 2470 BCE, since after this period the precession of the Earth's axis would have shifted this vertical line from the direction of true north. Spence mentioned the changing accuracy of the orientation as a calibrator for the period of pyramid construction. This orientation method was criticized by e.g.,

---

[19] Spence, K. 2000, *Nature* **408**, 320-324.
[20] Note various critiques of Spence's work based, at least partly, on her not correcting for the declination of the stars that puts her revision of the chronology in doubt.

Schaeffer[21], since limitations in naked-eye vision would have introduced by themselves large orientation errors. Moreover, it would have been useful only for a relatively short period because of the precession drift.

However, one wonders why was there a need to orient the pyramids using stars at night, when the North direction could be found reasonably well in daytime by using a shadow stick ("gnomon") or a similar method. This, from a personal point of view, seems much more likely, since the orientation to the cardinal points would have been performed in full light, when the construction work at the site probably took place.

The design of the pyramid of Khufu includes also a series of symbolic alignments in the ventilation channels emanating from two burial chambers in the interior of the pyramid. As a consequence of the general cardinal grid of the necropolis, the Sphinx, a personification of the god Horus at the horizon at least from the New Kingdom onwards, faces the equinoctial rising sun. However, the claim that the general pattern of the necropolis also encapsulates a series of additional topographic and astronomical alignments that create a cosmic landscape, reflecting the ancient Egyptian world-view, is debatable[22].

Alignments with astronomical meaning were searched for, and found, in numerous Egyptian monuments, temples and tombs. An extensive discussion of such alignments can be found in the series of papers by Belmonte and Shaltout. The conclusions of their series of expeditions, where various orientations were measured, were recently summarized[23]:

"(i) The temples of the Nile Valley and the Delta were orientated according to the Nile [. . .]
(ii) The temples were also astronomically oriented... This problem was solved by the selection of appropriate orientations of one or the other class at different sites so that they would be more or less compatible with the Nile course (quarter-cardinal directions are a good example of this), or by the deliberate election of selected places in Egypt where the Nile prescription and a conspicuous astronomical orientation were simultaneously achieved, as in the case of the temple of Karnak...
(iii) Among astronomical orientations, there were three, and only three, kinds of targets. One was probably related to different celestial configurations of the stars of Meskhetyu (Ursa Major) in order to get a near or accurate Meridian orientation. This primary axis could have been rotated later by an eighth, a quarter or half a circumference to obtain any possible cardinal or quarter-cardinal direction (families I, VI and VII). The second kind of targets had a markedly solar character and was fundamentally related to important time marks of the annual cycle and/or the civil calendar (families I, II, II? and III). Finally, the third group of targets was formed by the two brightest stars of the ancient Egyptian skies, Sirius and Canopus (families IV and V, respectively). These customs were present during most of Egyptian history and in the different areas of the country…and indicate an interest in the direction of solar orientation at sunrise and towards the North."

---

[21] For instance, as quoted in http://www.grahamhancock.com/forum/BauvalR4-p1.htm#ftn12.
[22] E.g., Krupp, E.C. at http://www.antiquityofman.com/Krupp_refutes_Bauval_and_Roy.html.
[23] Belmonte, J.A., Shaltout, M. 2010. *Keeping Ma'at: An astronomical approach to the orientation of the temples in ancient Egypt*. Advances in Space Research **46**, 532-539.

An interesting and original possibility that ancient peoples paid attention to a completely different astronomical phenomenon has been proposed by André Maucherat. He noticed the similarity in shape of the ancient Egyptian crowns with the cone made by the zodiacal light (ZL). Maucherat gives a number of examples where the inclination of the crown as depicted in murals seems to match the tilt of the ZL at the specific latitude of the site[24]. Most of the crowns shown by Maucherat in his book can be dated to the middle of the 2nd millennium BCE, with the exception of the much older Sekhemkhet crown shown in his Figure 19. Three "evening" crowns are shown there, one matching the inclination of the ZL cone at the spring equinox and two for the winter solstice, at the location where the site is located in Wadi Maghara (southwestern Sinai peninsula).

Maucherat discusses also the stele of Naram-Sin (or Naram-Suen; ~2254–2218 BCE), king of Akkad and grandson of Sargon, who predated by some five hundred years the Egyptian crowns whose inclinations he investigated. This stele, exhibited in the Louvre, celebrates Naram Sin's victory over king Satuni of the Lullubi people, and shows Naram-Sin having a godlike appearance near a conical shape representing a "hill" over which two "stars" are visible.

Maucherat found a match between the slope of the conical shape and the two bright objects over it with the ZL cone and the planets Jupiter and Venus as seen on 8 May 2221 BCE at 16:56 UT from a place at 34º32' N and 45º44E, which is near Qasr-e Shirin on the Iran-Iraq border, on the other hand. The location of the Qasr-e Shirin site, near a mountain pass at the foothills of the Zagros Mountains, matches the description of the mountainous homeland of the

---

[24] Maucherat, A. 2009, Les *Phenomènes Célestes et l'Art Égyptien*, Actilia Multimedia publishers, Theix, France.

Lullubi people. Since the dating of the Akkadian war against the Lullubi is extremely uncertain, Maucherat's fit could serve as a consistency check of Akkadian chronology.

A shallow survey of the archeoastronomical literature reveals therefore a number of Middle-Eastern artifacts dealing with orientation of monuments: the pyramids were astronomically oriented, either emphasizing the circumpolar northern stars or another celestial feature (perhaps the belt of Orion) on the eastern horizon. At least some temples were probably oriented to the East (perhaps to view the rising Sun at the equinoxes). Also, both the Egyptians and the Akkadians noted the Zodiacal Light and attached some importance to its inclination to the horizon. The question is whether there are other astronomically-oriented monuments or sites in Egypt that predate the pyramids, temples, etc. Apparently yes; at least one such site has been documented.

*Nabta Playa*[25]

The Nabta Playa is a site 800-km south of Cairo or about 100-km west of Abu Simbel in southern Egypt. It was investigated by McKim Malville et al.[26] who reported that it consists of a 4-m diameter circle of upright, narrow stone slabs forming a cromlech, sited on top of a small sandy knoll at the end of a shallow valley. The valley contains a number of stone mounds with remains of cattle and sheep sacrifices. The remains are dated from 5500 to 4900 BCE using radioactive carbon ($^{14}$C) dating.

The final Neolithic period at Nabta Playa began about 4600 BCE until the abandonment of the area about 3400 BCE. Within the cromlech, two pairs of stone slabs guide the eye: one oriented approximately to the north and the other to the position of the rising sun at the summer solstice; this would have been the sunrise point near the start of the rainy season. The site also contains long lines of monoliths that were investigated with ground-based GPS, satellite optical imagery, and satellite radar measurements[27]. The alignments of sandstone slabs are oriented to the rising points of bright stars (Sirius and Arcturus=α Bootis) on the Eastern horizon.

---

[25] A playa is a desert basin that periodically fills with water to form a temporary lake, which subsequently dries up. It is also known as an "alkali flat".

[26] McKim Malville, J., Schild, R., Wendorf, F., Brenmer, R. 2008. *Astronomy of Nabta Playa*. in African Cultural Astronomy-Current Archaeoastronomy and Ethnoastronomy Research in Africa J. Holbrook et al. (eds.), Springer Science+Business Media B.V., pp. 131-143.

[27] Rosen, P.A., Brophy, T., Shimada, M. 2008. *TerraSAR-X Spotlight Interferometric Observations of Archaeoastronomical Structures at Nabta Playa, Egypt*, poster paper presented at the TerraSAR-X Science Team Meeting - 25/26.11.2008.

If we accept the astronomical explanation for the orientation of the stones, then the megaliths of Nabta Playa indicate a very early interest in astronomy, more than 6000 years ago, on the part of a pre-agrarian society. However, it is possible to find something that appears much more astronomy-linked (although also more recent) that the stones of Nabta Playa. This is the Nebra disk which, despite is rather similar sounding name, does not originate in Africa at all.

*Nebra disk*

The Nebra disk is a bronze artifact found in 1989 by illegal treasure hunters in Germany, near Nebra (Unstrut), a small town in Saxony-Anhalt. The 32-cm diameter metal disk is considered authentic and dates from ~1600 BCE based on associated finds: two bronze swords, a chisel, axe heads, and bracelets typical for the 16$^{th}$ century BCE. In particular, a wood particle found adhering to one of the swords was carbon-dated to 1600-1550 BCE. As the disk was not excavated using accepted archaeological methods, it had to be authenticated from microphotography of the corrosion crystals; these produce images not reproducible by fakers. An initial analysis of trace elements by synchrotron radiation driven X-ray fluorescence[28] shows that the copper originated in Austria, while the gold and the tin content of the bronze were from Cornwall.

The disk was apparently developed in stages. The earliest version of the disk had thirty-two small round gold circles attached (of which only thirty remain visible), a large circular gold plate, and a large crescent-shaped gold plate. The circular plate may be interpreted as either the Sun or the full Moon, the crescent shape as the crescent Moon (or either the Sun or the Moon undergoing eclipse), and the dots as stars, with the cluster of seven dots likely representing the Pleiades. Later, two arcs (constructed from gold of a different origin as shown by its chemical impurities) were added at opposite edges of the disk. The two arcs span an angle of 82°, matching the angle between the horizon positions of sunrise and sunset at the summer and winter solstices for the 51°N latitude of the location where the disk was found. Given that the arcs seem to relate to solar phenomena, it is likely the circular gold plate represents the Sun, not the Moon. Finally, another arc was added at the bottom. This arc is called "the Sun boat" and is also made of gold, but from a different origin. When the disk was buried it had thirty-nine or forty holes, each about 3 mm in diameter, perforated around its perimeter.

---

[28] Pernicka, E., Radtke, M., Riesemeier, H., Wunderlich, C.-H. 2003, *European Network of Competence at 1600 BC*, BESSY – Highlights 2003 – Scientific Highligths, 8.

The question is what role the Nebra disk fulfilled: was it a religious artifact, or did it serve to instruct others as to the directions of sunrise and sunset at various seasons? Much has been made of the proximity of the site where the Nebra disk was found and the even older Goseck circle, some 20-km away. The discovery site identified by the disk finders is a prehistoric enclosure encircling the top of a 252-m elevation some 60-km west of Leipzig. The Goseck circle is a set of concentric ditches 75-m across with two palisade rings containing gates, and is presumed to have been a solar observatory. This is because two of the three openings have solar connections: one is directed to the sunrise and the other to the sunset points of the winter solstice (the renewal of the seasons) as seen from the center of the circle. Potsherds found at the site suggest that the Goseck observatory was built around 4900 BCE[29]. Goseck is one of the 250+ ring-ditches in Germany, Austria and Croatia identified by aerial surveys, with most of them not yet investigated and their astronomical connection is unclear.

As the Goseck circle and the Nebra sky disk do not date from the same era, a link between them has been speculated about, but remains entirely unproven. Similarly, the resemblance of the lower arc on the disk to the Egyptian Sun boats is not supported by evidence. However, both sites (accepting their astronomical explanations) seem to indicate that attention was given to celestial phenomena as far back as ~5000 BCE.

*The Antikythera mechanism*

If the Nebra disk served some cultist or educational purposes in the same period when the Minoan civilization reached its peak, then another artifact found not too far from Crete and dated considerably later, in the 2nd century BCE, is even more fascinating. This is the Antikythera mechanism discovered in 1901 among the remains of an ancient shipwreck in the Mediterranean Sea, half-way between Crete and the Peloponnesus. It was possibly looted from Mithridates' capital of Sinopi on the southern shore of the Black Sea[30]. Sinopi is a small city at the northernmost edge of the Turkish side of Black Sea coast. It was originally Hittite, but Greeks re-founded it in the 7th century BCE.

Even though only some bronze fragments were found, it is accepted that the Antikythera mechanism was a machine used to predict celestial appearances using 30 or more toothed wheels activated by one external handle. Although heavily corroded, and embedded in marine sediments, the mechanism has been studied in detail using extremely high resolution X-ray photographs that allowed also the reading of the Greek letters inscribed in the metal. Originally, the machine was enclosed in a wooden box roughly 30 by 20-cm. The front side had a large circular dial with two concentric scales and pointers. One scale, inscribed with names of the months, was divided into the 365 days of the Egyptian (or Sothic) year; the other, divided into 360 degrees, was marked with the 12 signs of the zodiac. The calendar dial could be turned to compensate for the extra one-quarter day of the solar year (of 365.2422 days) by turning the scale backwards one day every four years.

---

[29] Boser, U. 2006, *Solar Circle*, Archeology **59(4)**, abstract http://www.archaeology.org/0607/abstracts/henge.html.
[30] Marchant, J. 2010, *Mechanical Inspiration,* Nature **468**, 496-498.

The pointers on the front dial were probably showing the date and the position at the desired date of the Sun, Moon and probably the five planets known at the time, although the gears for these have not been found. A revolving, half black and half silver ball displayed the phase of the Moon for the date. The back of the device had two spiral dials with pointers, one above the other. The one at the top showed a 235-month calendar, probably because after 235 months (19 years) the distribution of new Moons in the solar year repeats itself; this is the Metonic cycle. The bottom spiral represented the 223-month eclipse cycle (Saros). These two cycles were originally derived by Babylonian astronomers.

The X-ray images revealed also that a mechanism within the device's gears modeled the varying motion of the Moon on its elliptical orbit around Earth, which is faster at some points in its orbit than at others. Apollonius of Perga[31] or Hipparchos[32] proposed in the 2$^{nd}$ century BCE that this variation in the Moon's motion could be explained by superimposing one circular orbit onto another having a different centre ('eccentric' or "epicyclic').

Intriguingly, mechanical gears as devices and astronomical epicycles as cosmological models seem to have appeared at about the same time, with the gears perhaps a little earlier. In the 3$^{rd}$ century BCE Archimedes used simple gears to change the size of an applied force. It is therefore possible that the astronomical epicycles of the Greek cosmology were not a philosophical innovation but a mechanical one, implying that technological development had, in this case, predated the theoretical work on cosmogology. It is also possible that the Antikythera mechanism was not unique, since Cicero refers to similar gear mechanisms made by Archimedes and by Posidonius.

We have seen here indications that early humans became interested in celestial phenomena, noted the cyclical occurrence of some events and their correlation with climatic phenomena that could impact their survival, and developed methods to use these celestial phenomena to "predict" or extrapolate when the future cyclical phenomena will occur. All these developments, as described above, took place in the European and Mediterranean regions from about 6000 BCE until close to the beginning of the Common Era. This begs the question whether other astronomically-related sites exit on other continents, and obviously the answer is, again, yes.

---

[31] Lived from about 262 to about 190 BCE.
[32] Lived from about 190 to about 120 BCE.

*Chankillo*

One could argue that the famous Nazca lines in Peru have astronomical meanings. However, Gerald Hawkins studied in 1973 the correlation of 186 Nazca lines with astronomical alignments and found that only 20% had any astronomical orientation; this could be expected by pure chance[33]. In 1982, Anthony Aveni obtained similar results[34], which could indicate a lack of astronomical interest in the prehistoric Americas. However, thanks to a discussion with Yvon Georgelin at the Observatoire Astronomique de Marseille-Provence, I learned of a location in South America that does seem to have served as a solar or lunar observatory; this is Chankillo in Southern Peru, studied by Ghezzi and Ruggles[35] and dated to about 300 BCE, based on radioactive carbon dating. The site appears to have been used until the Spanish Conquest.

Briefly, Chankillo consists of two features: a 300-m-long hilltop structure built in a remote location and heavily fortified with massive walls, restricting gates, and a parapet called the "fort", and about 800-m away to the South-East the "Thirteen Towers", a row of 13 cuboid constructions between 2 and 6-m tall and spaced from 4.7 to 5.1-m, placed on an approximately North-South line some 200-m long on the ridge of a low hill. Ghezzi & Ruggles identified two structures, one to the East of the 13 towers and the other to the West (in the fort), as possible observation points for the sunrise and sunset positions at certain dates such as the solstices and the equinox, as well as the date of the zenith passage of the Sun.

Alternatively, Georgelin[36] proposed that the site was used to observe the rising of the Moon from the "fort", and the 13 towers served to account for the changing declination of the Moon on all positions between the north and south major standstills. The primary function of the site was, according to Georgelin, to enable the prediction of times when a Solar or Lunar eclipse would be possible.

---

[33] Hawkins, G. 1973. *Astronomical alinements in Britain, Egypt and Peru.* Phil.Trans.R.soc.Lond. A. **276**, 157-167.
[34] Aveni, A. F. 2000. *Between the Lines: The Mystery of the Giant Ground Drawings of Ancient Nasca, Peru* . Austin Texas: University of Texas Press.
[35] Ghezzi, I. and Ruggles, C. 2007. *Chankillo: A 2300-Year-Old Solar Observatory in Southern Peru*, Science, **315**, 1239-1243.
[36] Private communication to NB.

It makes reasonably good sense to accept Chankillo as a genuine astronomical observatory in the 3$^{rd}$ century BCE, irrespective of whether it was sunrises and sunsets that were observed as argued by Ghezzi and Ruggles, or only moonrises as seen from the "fort" as Georgelin thinks. This is the place to ask similar questions, about the existence of astronomically-related sites, in Asia and, as shown for the Americas, the answer to this is also yes, such sites do exist.

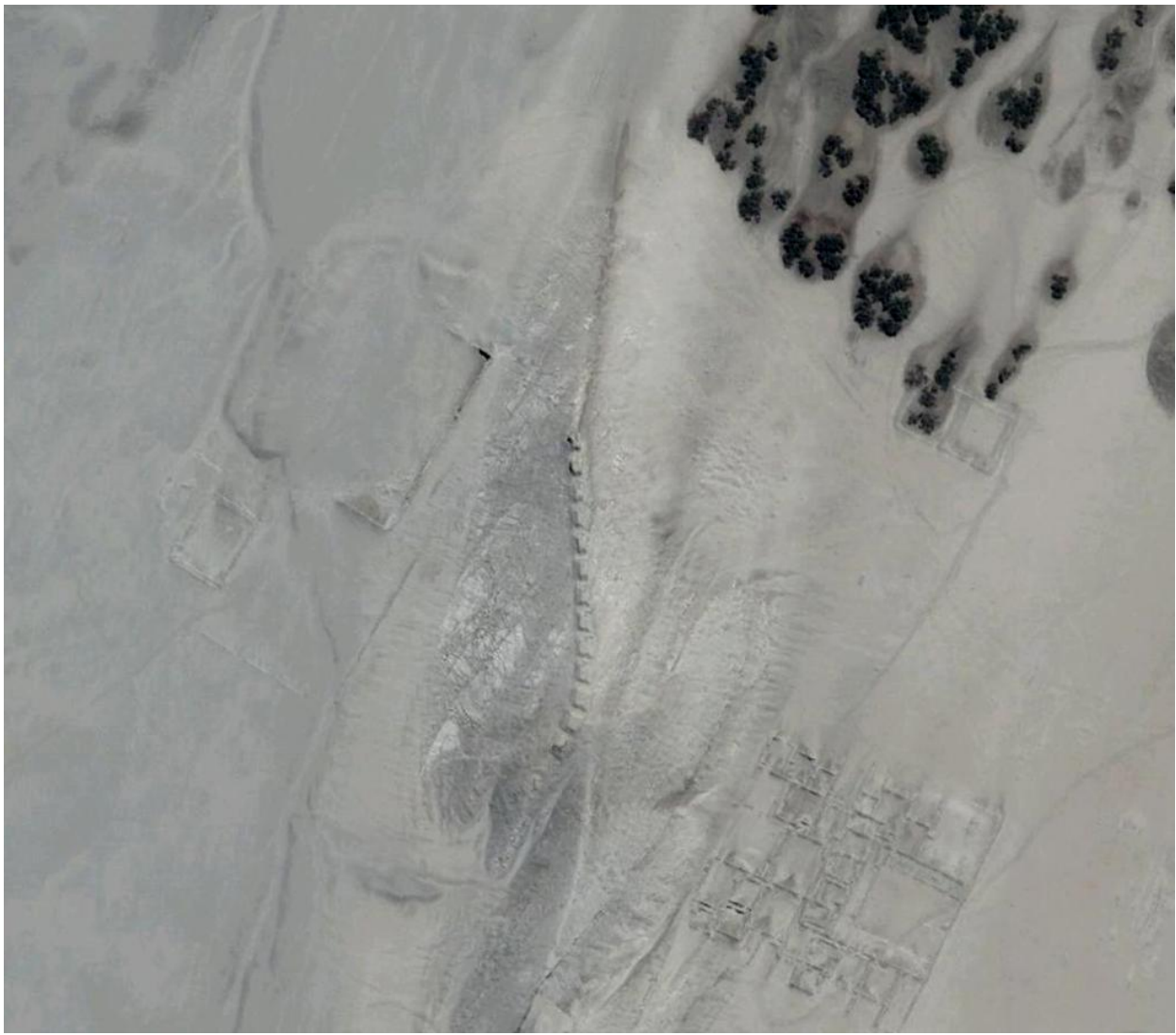

Figure 2: Chankillo; the 13 towers at the center of the image (from Google Earth). North at top and East at right.

*Korean megaliths*

Archaeological data indicate that the Korean Peninsula and its vicinity was inhabited more than 500000 years ago during the Paleolithic and Neolithic periods. At the end of the last Ice Age, about 10000 years ago, the Korean Peninsula became part of the temperate zone showing the climatic changes of the four seasons. During this period the Neolithic people who settled on the Korean Peninsula and its contiguous areas started manufacturing polished stone implements and coarse earthenware, and building pit dwellings that formed villages.

The techniques of hunting, gathering and fishing developed further and along with farming. Ancient Chinese historical texts refer to an ethnic group named "Yemaek" that established itself on the northern Korean Peninsula and Manchuria, engaging in farming and animal husbandry. These people, different from the nomadic people inhabiting the grasslands to the north of the Great Wall of China, were later assimilated into the Korean people.

The early Go-Joseon (or Ko Chosǒn) era[37] in Korean history is often linked to the ~50000 megalithic dolmens distributed over the entire Korean Peninsula and Manchuria, which often served as tombs and are rarely found at Chinese archaeological sites. With the help of my former graduate student, Dr. Young-Jun Choi from the Korea Astronomy and Space Science Institute, I obtained a copy of the PhD thesis of Dr. H.-J. Yang that deals with this topic[38].

Chapter 4 of Yang's thesis deals with astronomical aspects of the Korean dolmens, which date from the 1st millennium BCE (7th to the 3rd centuries BCE; in the Go-Joseon period). While Yang could not identify specific directions for the long axes of the dolmens except those related to the local topography (some cover stones, located on mountain ridges, were generally placed parallel to the ridge direction), he identified patterns of holes called "cup-marks" in some of the dolmen cover stones that seem to represent star maps, in particular the constellations of Ursa Major and Sagittarius. The Korean dolmen cup-marks were generally carved on the cover stone in a South-Easterly direction, regardless of the direction of the long axes of the cover stones[39]; this orientation was interpreted as attention being paid to the direction of sunrise.

---

[37] Founded by Tan-gun on about 2,333 BCE.
[38] Yang, H-J., 2004, *Analysis of Korean Historical Astronomical Records*, PhD thesis at The Graduate School, Kyungpook National University, Korea.
[39] Hong-Jin Yang, Changbom Park, Myeong-Gu Park, 2010. *Study on the dolmens with cup-marks in South Korea*, Journal of Korean Petroglyph, **13**, 1-15 (in Korean, English abstract).

Yang does not show a correlation analysis between the patterns of the cup-holes and the celestial constellations, but his argument is supported by the presence of recognizable sky maps on later (5th century CE) Goguryeo, or Koguryo, Korean tombs. The earliest Korean artifact depicting a stellar map is a painted lacquered box found in the Zeng Hou Yi tomb in Habei (burial from about 433 BCE). This map shows the 28 constellations known to Koreans at that time. These show that the early inhabitants of Korea also showed interest in the sky, although theirs was less inclined toward the Sun and more to the stars.

*Interlude*

After showing that it is possible to find astronomically-related sites on Europe, Africa, Asia and the Americas, it is possible to combine all the information presented so far to see whether the four initial questions have been answered. I asked first:

(a) What kinds of artifacts can be associated with archeoastronomy?
(b) What dates can one assign to the various artifacts?
(c) How much effort was required to produce the artifact?
(d) Does this interpretation make sense?

I have shown that many types of artifacts have been associated with archeoastronomy. These range from monumental structures such as Egyptian temples, megalithic sites from Korea through Europe and Africa to Peru, to sophisticated instruments used to predict the occurrence of relevant celestial events (Chankillo observatory and Antikythera mechanism). The dating of the various artifacts shows that the interest in archeoastronomy started in about the 6th millennium BCE, probably in the south part of Egypt. European sites, such as Stonehenge and the earlier Goseck circle, appear later on the world stage that the Nabta Playa alignments. One immediate conclusion is that there does not seem to be any synchronicity between the appearances of cromlechs at various sites. The question of whether astronomically-related knowledge diffused from a single location throughout the world, or appeared independently on all sites though at different times, is not discussed here.

Some sites required very significant labor efforts to design and erect; others were much more modest and could have been completed by a few individuals. For instance, some calculations indicate that some 13000 people were employed for about 10 years to build the Khufu pyramid, for a total cost of 111 million jugs of beer and 126 million loafs of bread[40]. I tried, unsuccessfully, to search for an agreed size of the Egyptian population in the same period; Smith asserts that the Egypt population was then between one and 1.5 million people. Two other indications show that the entire population was of order one million: Nekhen[41], the early capital of Upper Egypt, had at most 10000 inhabitants around 3100 BCE, and Memphis, the new capital of Ancient Egypt after Nekhen, which was founded in that period, had between 6000 and 30000 inhabitants by 2250 BCE. It seems that the pyramid construction could have been accommodated by the Egyptian economy if it would have been conceived of as a "national project".

But could the astronomical orientation accuracy, requiring highly-trained and dedicated individuals, be achieved? How many specialists would have to drawn from a population of about one million to achieve this accuracy, on top of the other astronomical tasks such as observing the heliacal rising of Sirius? How long did they train to perform their duties without contributing to the tribe's or nation's survival? This question has not been dealt with in the literature, as far as I could find, and neither did I find a quick answer to the ratio of shamans/witch doctors/medicine men and women to the size of the population they served. An analogy can be adopted from the case of the present American Indian Pawnee nation[42]. A report deplores the dwindling number of medicine men, which results in a serious reduction of ceremonies, and discusses the present corpus of medicine men:

**Among the Pawnees** "`the medicine men ranked socially next to the chiefs and priests… In every Indian tribe there were a number of persons, called medicine men by the whites, who were regarded as the possessors of supernatural powers which enabled them to recognize and cure disease.`"[43]

Since the Pawnees live in villages of about 300 people, an upper limit to the sought-after ratio might be one medicine person for every 300 persons of this nation. If this proportion was similar in ancient Egypt, for example, then from a population of 1.5 million one could expect some 5000 "wise men" and out of those perhaps a very small number of scientific leaders of their generation.

---

[40] Smith, C.B. 1999, *Program Management B.C.* Civil engineering magazine; on-line at http://web.archive.org/web/20070608101037/http://www.pubs.asce.org/ceonline/0699feat.html.
[41] The Narmer palette was found at this site.
[42] Bill Donovan 1999. *NAVAJO NATION STRUGGLES TO PRESERVE MEDICINE-MAN TRADITION*. Web entry at http://www.mail-archive.com/nativenews@mlists.net/msg00855.html retrieved 26 March 2011.
[43] http://www.archive.org/stream/annualceremonyof08lint/annualceremonyof08lint_djvu.txt for a 1923 report on the Annual Ceremony of the Pawnee Medicine Men kept at the University of Illinois.

After showing that the synchronicity argument for the construction of the cromlechs is not tenable, it is possible to put the archeoastronomy sites discussed here on a geographic and timeline plane. This shows what was already stated; the "epidemic" of astronomical inventions and constructions started in the 7-6th millennia BCE somewhere in Africa by people who built stone circles, and one-two millennia later in Europe dug circular trenches and built wooden palisades with the same shape. I wanted to see whether "stone circles" existed even before the Nabta Playa or the Goseck circle and, to my amazement, I found that yes, they did exist.

*Göbekli Tepe*

Göbekli Tepe ("Potbelly Hill", in Turkish) is a site in the south-east Turkey, some 30-km from the Syrian border, at the Fertile Crescent's northernmost tip. It was discovered in 1964, but archeological investigations started only in 1994 by a German team headed by Klaus Schmidt[44]. The site is a collection of megaliths arranged in circles, the largest being some 30-m in diameter. The stones are up to 5-m tall and some are incised with rather complicated petroglyphs depicting animals (ibis birds, wild boars, lizards, etc.). The site was active from about 11000 BCE to about 8000 BCE, when it was completely and deliberately buried and remained buried until discovered. The earliest carbon-dated sample from the site is from 9130–8800 BCE.

Göbekli Tepe is close to the site where wheat was first domesticated; DNA analysis of modern domesticated wheat compared with that of wild wheat strains show that the modern wheat's DNA is closest in structure to that of the wild wheat found on Mount Karaca Dağ, some 30-km from Göbekli Tepe[45]. Schmidt speculated that the early hunter-gatherer society transformed into an agricultural one at this site, starting with the organized exploitation of wild crops followed by deliberate cultivation.

So far, four structures consisting of T-shaped pillars with partial stacked-rock walls, with diameters between 10 and 30-m, have been uncovered at Göbekli Tepe, but geophysical surveys indicate the existence of at least 16 additional stone circles or elliptical ovals. It is likely that the site served for cultist purposes, since no signs of regular continuous habitation were found. Schmidt and colleagues estimate that up to 500 persons were required to extract the 10–20 ton pillars (some weighing as much as 50 tons) from local quarries and to move them a few hundred meters to the site[46].

---

[44] Schmidt, K. 2000. *Göbekli Tepe, Southeastern Turkey. A Preliminary Report on the 1995-1999 Excavations.* Paleorient **26(1)**, 45-54.
[45] Heun, M., Schäfer-Pregl, R., Klawan, D., Castagna, R., Accerbi, M., Borghi, B., Salamini, F. 1997. *Site of Einkorn Wheat Domestication Identified by DNA Fingerprinting*, Science **278**, 1312-1314.

While the structures may have been primarily temples, smaller domestic buildings have been uncovered more recently. Schmidt believes that the site was a pilgrimage destination attracting distant worshipers, and may have served also as a cemetery[47]. Butchered bones found in large numbers from local game, such as deer, gazelle, pigs, and geese, have been identified as refuse derived from hunting and food prepared for the congregants. The site, as already mentioned, was deliberately backfilled sometime after 8000 BCE: the buildings are covered with settlement refuse that must have been brought from elsewhere.

I must stress that in most publications or items related to the Göbekli Tepe site I saw there was not any mention of a possible astronomical link. In fact, it is not even clear whether the round structures were originally covered, or which way the openings were facing[48], whether the horizon was visible from any of the openings, etc. However, the similarity of the Göbekli Tepe stone circles with other megalithic structures, later in time and distant in geography, makes one wonder. The only exceptions claiming a possible Göbekli Tepe astronomical connection of which I am aware are in a contribution by Belmonte and González García[49] where they identify some astronomical motifs (a crescent Moon and a disk, the Scorpio and Leo constellations) on Göbekli Tepe pillars and report that one of the structures is "almost perfectly aligned according to the four cardinal points" (see also Belmonte[50]), indicating some astronomical interest as old as 11000 BCE from the part of a proto-agrarian culture.

*The Jericho tower*

An impressive megalithic structure is embedded in the walls of ancient Jericho and was uncovered in the 1950s by Kathleen Kenyon; this is an ancient tower at least 8-m high (the top is missing) and with a 9-m lower diameter containing a 20-step internal staircase that predates by thousands of years the oldest Egyptian stone structures. Barkai and Liran[51] argued that the Jericho tower, dated from about 8300 BCE, was built to provide an alignment with the peak of the Quruntul Mountain, approximately 1300-m away and about 350-m taller than the Jericho terrain, in the direction of the setting Sun at the summer solstice, and that the internal staircase is aligned in the same direction.

---

[46] Scham, S. 2008. *The World's First Temple*, Archeology **61**(6).
[47] Curry, A. 2008. *Gobekli Tepe: The World's First Temple?* Smithsonian on-line magazine at http://www.smithsonianmag.com/history-archaeology/gobekli-tepe.html?c=y&page=1.
[48] Dietrich, O. (the Deutsches Archäologisches Institut Orient-Abteilung), private communication to NB on 17 January 2011.
[49] Belmonte, J.A. and González García, A.C. 2010. *Astronomy, landscape and power in Eastern Anatolia.* Contribution at SEAC 2010.
[50] Belmonte, J. A. 2010. *Finding Our Place in the Cosmos: The Role of Astronomy In Ancient Cultures*. Journal of Cosmology **9**, 2052-2062.

The alignment of a staircase with an external backsight is similar to that of the dromos (passage, with stairs) in Tomb 165 at the late Minoan II-IIIB period at Armeno in Crete, which is dated from 1450 to 1190 BCE. That dromos is viewing the peak of Mt. Vrysinas, as shown in the Hoskin's book[52].

The Jericho tower is a structure built as part of the "pre-pottery Neolithic A" town of Jericho. Barkai and Liran argue that the explanation of the tower as fortification is not tenable since no likely enemies were identified. An alternative explanation, that the tower served to deflect flood waters, is also not tenable since no significant water sources were located and also since this does not explain the tower height.

Barkai and Liran proposed that the tower was constructed at its specific location in order to serve as a backsight, allowing the inhabitants of ancient Jericho to see the Sun setting behind the peak of the Quruntul Mountain at the summer solstice. While the authors claim that Midsummer in 8300 BCE would have happened in contemporary late February or early March, this is not confirmed by calculations with accurate astronomical programs. For the coordinates of Jericho, the longest daylight period in 8300 BCE, which corresponds to Midsummer, would have been August 19. On this date, and considering the ~15º horizon elevation difference with the Quruntul being higher, sunset[53] would have taken place at 17:50 with a solar azimuth of 289º. This is almost exactly the azimuth of the internal staircase of the tower and the direction to the mountain peak. The Jericho tower can thus be considered another astronomical megalith constructed about 1000-km south of Göbekli Tepe, just about when the use of Göbekli Tepe was terminated and that site was deliberately buried.

---

[51] Barkai, R. and Liran, R. 2008. *Midsummer Sunset at Neolithic Jericho*. Time and mind 1(3), 273-284.
[52] Hoskin, M. 2001. *Tombs, Temples and their orientation*. Ocarina Books, UK, p. 221
[53] Calculated using the software Carte du Ciel v3.0.

*Skeptical remarks and some hope*

Despite the abundant indications in the literature since archeoastronomy became a valid topic of scientific interest, I worry whether the different investigators do not try to read too much in the findings thus over-interpreting the archeological evidence.

David Dearborn remarked during an interview[54] on the astronomical interpretations of artifacts that "…a well known example of a possibly astronomical image is the petroglyph showing what appears to be a crescent moon and with a large spot nearby. The organization of the two elements is consistent with it being a relatively realistic representation of the 1054 supernova. Visible in the daytime sky, this object was certain to be of interest to anyone familiar with the sky. The uncertainty of date and local uniqueness make it difficult to prove this interpretation…Astronomical "explanations" can be fitted notoriously easily to preexisting alignments.... Fortuitous stellar alignments are particularly likely, given the number of stars in the sky and the fact that their positions change steadily over the centuries owing to precession."

It would help having more statistical support for the various proposals. For instance, while Belmonte and Shaltout present a rose-of-the-winds plot of sight-line azimuths, they do not show what the statistical significance of this plot is. This is, more or less, what Hoskins did when presenting the results of his extensive investigations of Mediterranean tombs and temples[55]. It is a pity that Maucherat's original suggestion, of the alignment of Egyptian crown with the local zodiacal light cone, which is very interesting, is not supported statistically at this time. It would be helpful if a complete survey of images of ancient (and dated) Egyptian crowns with their location and inclination measured, followed by a check on the correlation of their inclinations with the expected ZL inclinations for the crowns' sites.

Regarding the Nabta Playa, the 2007 paper by McKim Malville et al. in African Skies tried to present a more scientific approach by quoting uncertainties in the azimuth lines, but did not offer a statistical assessment of whether their explanation holds only for the stars they list, or whether other celestial bodies might also be implied. These examples show that a more rigorous approach is needed, probably to be combined with advanced research methods. This has been stressed as far back as 1968 (cf. Hawkins, op.cit.). However, some recent papers contain more rigorous and quantitative treatments, as well as the introducing supporting evidence.

---

[54] On-line at http://archaeology.about.com/cs/archaeoastronomy/a/dearborn_2.htm.
[55] Hoskins, M. 2001. *Tombs, Temples And Their Orientations*. Ocarina Books.

One example is the study of Langebroek[56] of the orientation of dolmens in the Netherlands and in nearby Germany. These dolmens belong to the Funnelbeaker culture[57], are dated from around 3400-3000 BCE, and show a preferred orientation not explainable by the orientation to the landscape (main direction of ridges/slopes, etc.), as shown by the Korean dolmens described above. There are a few exceptions, where the dolmens are aligned with an ancient road. The vast majority of the dolmens, some 38 of the 43 investigated, show an azimuth distribution limited by the minor lunar standstill rise positions; these are the azimuths of Moonrise when the Moon changes its declination from +18.5° to −18.5° within two weeks. Langebroek proposed that this moonrise orientation between the northernmost midsummer rise and the southernmost midwinter rise point is a consequence of orientations on the Full Moon rise points near the autumn equinox (harvest time), and not specifically to standstill points.

The recent work by Efrosini Boutsikas on the orientation of Greek temples[58] is worth of special attention since she not only shows results for a large, statistically-significant sanctuary sample (N=107), but also connects in her works the astronomical orientations with cultural inferences derived from written evidence. This has to do, e.g., with the cult of Artemis Orthia in Sparta[59] where the choice of the temple location was selected so that the Procession of the Girls with their offerings to the goddess could take place near the heliacal rising of the Pleiades star cluster. Her paper with Salt[60] discussing the location of the Delphi temple of Apollo (where the famous oracle resided) argues that the heliacal rising of the constellation Delphinus (described already by Aratus in the 3$^{rd}$ century BCE) served as annual marker for the pilgrimage to Delphi. Because of the mountains surrounding the site, the constellation rose heliacally about a month later at the temple's location in Delphi than elsewhere in Greece, allowing sufficient time to prospective pilgrims to travel to the oracle.

---

[56] Langebroek, M. 1999. *Huilen naar de Maan.* PIT **1(2),** 8-13.
[57] The principal North Central European megalithic culture of late Neolithic Europe (from about 4000 to about 2700 BCE).
[58] Boutsikas, E. 2007-8. Placing Greek Temples: *An Archaeoastronomical Study of the Orientation of Ancient Greek Religious Structures*. Archeoastronomy **XXI**, 4-19.
[59] Boutsikas, E. and Ruggles, C. 2011. *Temples, Stars, and Ritual Landscapes: The Potential for Archaeoastronomy in Ancient Greece*. American Journal of Archaeology **115**, 55-68.
[60] Salt, A. and Boutsikas, E. 2005. *Knowing when to consult the oracle at Delphi.* Antiquity **79**, 564–572.

*Even more ancient attention to celestial phenomena?*

Some of the more amazing pointers to very ancient astronomical interest have been published. One is the Taï (or Thaïs) bone, a part of a bovine rib measuring 87 mm × 27 mm, engraved on both faces and with one of its faces almost entirely covered by six rows of tiny parallel incisions[61], dated at 12800±300 BCE. Alexander Marshack examined microscopically each scratch mark and concluded that they recorded not only lunar observations, but also solar solstitial events. According to Marshak, the Taï plaque reveals a sequence of short horizontal lines or sections that carry irregular subsets of marks and reverse direction in each subsequent line.

Marshack's microscopic analysis showed that the sets of marks were made using various tools at different times and that the sequence could be characterized by clustering, variation, and periodicity. Marshak's assumption was that the sequence of lines and subsets of marks were a non-arithmetic form of lunar/solar observational recording. He based this conclusion on the binning of the notches in groups of 29, corresponding to the number of days between successive lunations.

Marshak suggested that the Taï bone records day-by-day lunar and solar observations undertaken over a time period as long as 3½ years. The markings appear to record the changing appearance of the Moon and, in particular, its crescent phases and times of invisibility. The shape of the overall pattern suggested to Marshak that the sequence was kept in step with the seasons by observations of the solstices. Marshak concluded that "…the Taï plaque and similar artifacts reflect a non-arithmetic observational astronomical skill and lore which lasted for 20,000 years".

Marshak's interpretation was criticized by d'Errico[62], who manufactured scratches on stone and bone and also used microscope inspection to investigate these. d'Errico claimed that the marks on the Taï bone could have been made almost simultaneously, therefore with the same tool and same hand, and may therefore represent no astronomical notation or long-duration record. The jury is still out on the astronomical interpretation of the Taï bone, but the International

---

[61] Ruggles, C. and Cotte, M. 2010. *Heritage Sites of Astronomy and Archaeoastronomy in the context of the UNESCO World Heritage Convention*. ICOMOS and IAU, ISBN 978-2-918086-01-7 (e-book).
[62] d'Errico, F. 1989. *Palaeolithic lunar calendars: a case of wishful thinking?* Current Anthropology 30(1): 117-118.

Astronomical Union has already adopted it as the first case in the Thematic Study of Heritage Sites[63]

An exciting possibility of even more ancient evidence for astronomical interest was suggested by Gore[64] in a 1903 letter to the Observatory journal. It was based on one of the Arabic names of Sirius: *al-schira al-abur* (Sirius which has passed across). Gore linked this name with Al Sufi's mention of a mythological explanation that Sirius crossed the Milky Way from North-east to South-West going in the direction of Canopus. This obviously happened during the Stone Age and, if correctly representing the human memory of an astronomical event, is truly amazing. The track of Sirius among the stars as seen from the Solar System, calculated by Rob van Gent and shown in Brosch (2008)[65], indicates that the Milky Way crossing by Sirius started some 50000 to 25000 years ago, about twice the age of the Taï bone. If this proposition is correct, it seems to indicate not only an interest in the appearance of the sky 50000 years ago, but also an ability to map the relative positions of stars and transmit this information throughout the generations. However, this possibility cannot be debated in the present paper since no astronomically-related artifact is involved.

---

[63] Ruggles, C. and Cotte, M. 2010. *Heritage Sites of Astronomy and Archaeoastronomy in the context of the UNESCO World Heritage Convention*. ICOMOS and IAU, ISBN 978-2-918086-01-7 (e-book), 16-18.
[64] Gore, J.E. 1903, *The proper motion of Sirius*, The Observatory, **26**, 391-392.
[65] Brosch, N. 2008. *Sirius Matters*. Springer. p. 58.

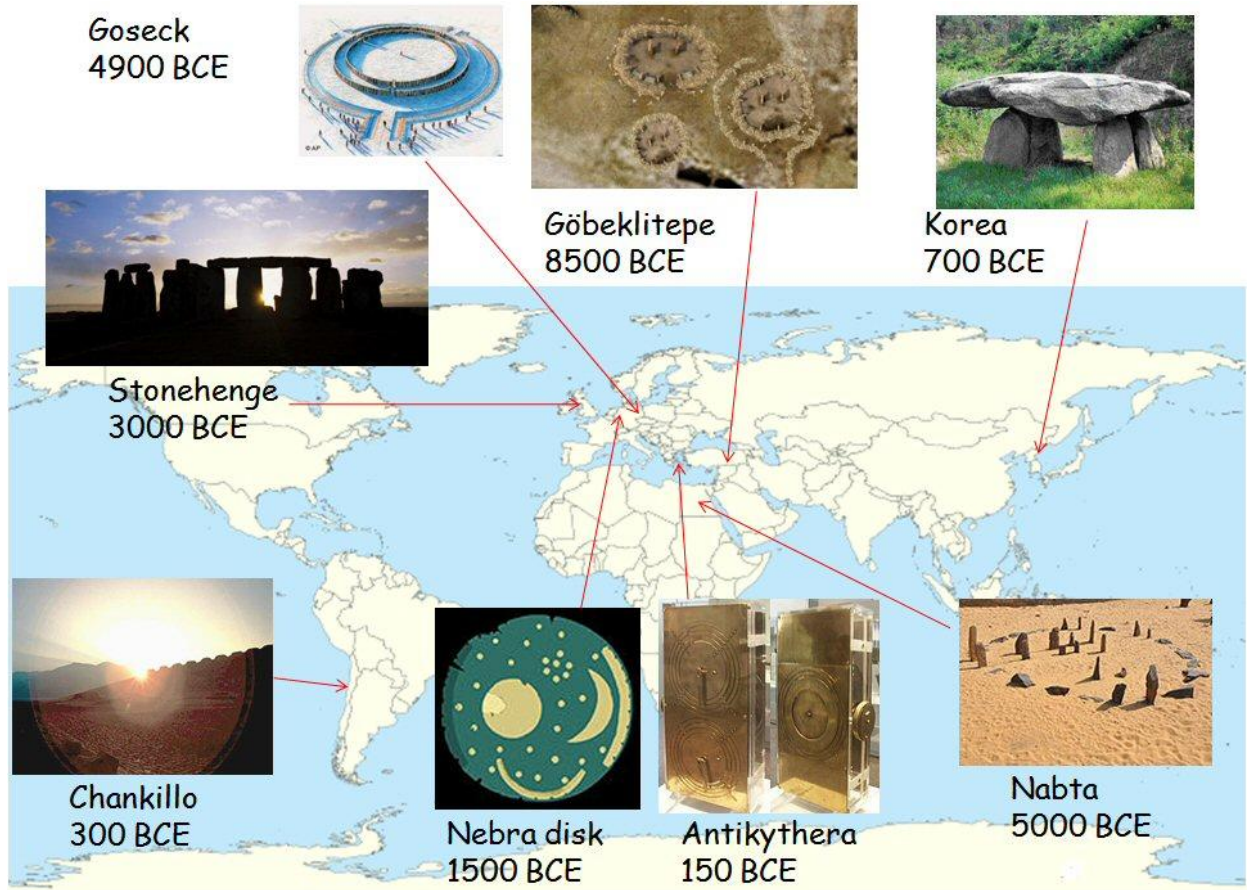

Figure 3: Geographical and temporal distribution of various artifacts and megalithic structures discussed here.

*Conclusions*

I have shown here that the chronology of the various sites connected with astronomy does not support an idea of synchronicity; similar cromlech sites appeared as far in the past as ~5000 BCE until as late as the beginning of the Common Era. On the other hand, the geographical distribution of the site combined with their dating seems to indicate an origin in the Middle East, presumably connected with the Fertile Crescent (see Figure 3). The various sites did not require extraordinary construction efforts; even the erection of the Great Pyramid, by far the most labor-intensive edifice, could conceivably be accommodated by the population of Egypt at that time if the project would have received a high national priority. A similar conclusion was reached by Atkinson (op. cit., p. 128) regarding the Neolithic mound from 2750 BCE at Silbury Hill; this could have been constructed by a team of 500 persons working for 15 years. Atkinson compared the intensity of this effort relative to the population size as being equivalent to the US space program in the early 1970s. Regarding the availability of qualified experts, an analogy with present-day Pawnee Native American Indians indicates that the required corpus of experts would not have been extremely large, but rather well within the capabilities of the different ancient societies. One conclusion that can be drawn from the phenomena studied or extrapolated from observations with artifacts such as those described here is that these are cyclical, not random, indicating that the early astronomers were looking for patterns and using these to predict future events.

On the research point of view, the examples presented here show that modern archeoastronomers do not invoke only astronomical alignments and directions, but attempt to support their field findings with written evidence, in the case of literate societies.

*Acknowledgements*

This paper resulted from a contribution to the archeoastronomy workshop held at IMéRA, Marseille on November 30, 2010. I acknowledge a residency at IMéRA in October and November 2010 when my interest in archeoastronomy was kindled. I am grateful for remarks of various draft stages of this paper received from Virginia Trimble, Hong-Jin Yang and Juan Belmonte. I am grateful to Rob van Gent for calculating the perceived motion of Sirius. Prof. Dr. Klaus Schmidt and Mr. Oliver Dietrich from the DFG Projekt Urfa at the Deutsches Archäologisches Institut are thanked for remarking on their Göbekli Tepe findings. Marco Langebroek kindly directed me to a copy of his paper on Dutch megaliths and offered additional explanations from his work in progress.